# Multi-modal Graph Neural Network for Early Diagnosis of Alzheimer's Disease from sMRI and PET Scans

Yanteng Zhang[a,b], Xiaohai He[a], Yi Hao Chan[b], Qizhi Teng[a], Jagath C. Rajapakse[b,*]

[a] *College of Electronics and Information Engineering, Sichuan University, Chengdu, 610065, China*
[b] *School of Computer Science and Engineering, Nanyang Technological University, Singapore 639798, Singapore*

**Abstract**

In recent years, deep learning models have been applied to neuroimaging data for early diagnosis of Alzheimer's disease (AD). Structural magnetic resonance imaging (sMRI) and positron emission tomography (PET) images provide structural and functional information about the brain, respectively. Combining these features leads to improved performance than using a single modality alone in building predictive models for AD diagnosis. However, current multi-modal approaches in deep learning, based on sMRI and PET, are mostly limited to convolutional neural networks, which do not facilitate integration of both image and phenotypic information of subjects. We propose to use graph neural networks (GNN) that are designed to deal with problems in non-Euclidean domains. In this study, we demonstrate how brain networks can be created from sMRI or PET images and be used in a population graph framework that can combine phenotypic information with imaging features of these brain networks. Then, we present a multi-modal GNN framework where each modality has its own branch of GNN and a technique is proposed to combine the multi-modal data at both the level of node vectors and adjacency matrices. Finally, we perform late fusion to combine the preliminary decisions made in each branch and produce a final prediction. As multi-modality data becomes available, multi-source and multi-modal is the trend of AD diagnosis. We conducted explorative experiments based on multi-modal imaging data combined with non-imaging phenotypic information for AD diagnosis and analyzed the impact of phenotypic information on diagnostic performance. Results from experiments demonstrated that our proposed multi-modal approach improves performance for AD diagnosis, and this study also provides technical reference and support the need for multivariate multi-modal diagnosis methods.

## 1. Introduction

Alzheimer's disease (AD) is a degenerative disease of the central nervous system, largely manifested in the form of memory, language, cognition, and even emotional disorders. The state between normal control (NC) and AD is called mild cognitive impairment (MCI) and more than half of MCI cases progress to AD [1]. Still, no cure or preventive drugs have been successfully developed for AD but early diagnosis of AD allows for early intervention measures that could delay the progression of the disease [2]. Therefore, early diagnosis and treatment of AD is of great significance to patients. At present, the clinical diagnosis of AD largely depends on a wide range of sources, including medical history, neurological assessments, behavioral tests, neuroimaging

---

[*] Corresponding author. E-mail: asjagath@ntu.edu.sg (Jagath C. Rajapakse).

scans, etc. [3]

Neuroimaging plays an important role in the identification of treatable causes of dementia and provides a stronger basis for the screening and early diagnosis of AD [4]. A variety of imaging methods including structural magnetic resonance imaging (sMRI) and positron emission tomography (PET) techniques for clinical image-assisted diagnosis, which provides information about brain structure and function, respectively. sMRI helps us to understand changes in brain structure features (such as volume and shape) and can be used to predict AD progression [5]. On the other hand, 18-fluorodeoxyglucose PET (FDG-PET) is a molecular diagnostic method to visualize glucose metabolism. The functional analysis of PET image is carried out by studying the degree of glucose metabolism [6]. Both structural and functional information are important when studying biological systems. Studies performed on only one modality is unable to capture both structural and functional aspects of the brain. To make up for these shortcomings, several studies [7] have used multi-modal methods to enhance features, leading to better prediction performance.

In recent years, deep learning methods have been proposed for the analysis and diagnosis of diseases related to cognitive impairment [8]. Although convolutional neural networks (CNN) can learn image representations effectively, they do not fully consider the correlation between the subjects. Furthermore, CNN provides limited extensibility for the integration of multi-modal datasets; one major downside is the need for all inputs to have the same dimensions if each channel represents a different modality. On the other hand, graph neural networks (GNN), which extends classical CNN to non-Euclidean space by using graph topology or feature propagation between neighborhood nodes [9], afford greater flexibility for multi-modal integration. Graph convolutional network (GCN) is a type of GNN that works directly on graphs and take advantage of relational information encoded in the graph structure [10]. For instance, when nodes are used to represent subjects (such as patients or healthy people), edges of the graph can store information about the similarity between nodes. GCNs can perform signal filtering and aggregate information from neighboring nodes to obtain improved feature representations, which in turn can be used for disease prediction and graph analysis [11][12]. The flexibility of choosing various combinations of data modalities for node vectors and adjacency matrices make GCN an ideal method to combine multi-modal images.

GNN methods have achieved commendable performance for AD diagnosis. Ktena et al. [13] used graph similarity measures between brain connectivity maps of functional MRI (fMRI) to construct a multi-layer GCN filter for AD prediction. Song et al. [14] constructed a multi-class GCN classifier based on structural connectivity, performed multi-class disease classification of four disease stages across the AD spectrum, and verified that the GCN classifier outperforms the SVM on a disease prediction task. Zhang et al. [15] proposed a GCN using multi-modal brain networks from various diffusion weighted imaging sequences to predict clinical indicators and verified the effectiveness of integrated multi-modal brain network in prediction tasks. Overall, these studies have shown the effectiveness of GCN in the diagnosis of brain related diseases.

In the above works that relied on GCN, fMRI and diffusion tensor imaging (DTI) data are usually used for graph analysis tasks [16][17]. There are established methods to construct individual brain networks from these modalities but there is no clear way to do so directly by regions-of-interest (ROI) features for modalities such as sMRI and PET (which are 3-dimensional, as compared to 4-dimensional fMRI and DTI data). This makes it challenging to construct a GCN based on sMRI and PET images. To solve this problem, we adopt a method of generating brain networks [18] via brain ROI features to obtain individual features of subjects. Then, we draw inspiration from the flexibility of graph-based analysis by combining the use of graph nodes to represent the individual features of subjects with the use of a sparse population matrix built using phenotypic information. Finally, a population-based GNN is constructed for the early diagnosis of AD based on sMRI and PET images via the multi-modal GNN framework. We propose combining of sMRI and PET information both at the level of node vectors as well as at the adjacency matrices. We show that our proposed approach led to improvements in model performance for both AD detection and prediction of sMCI versus pMCI. Furthermore, we perform ablation studies on the demographic features used and found that combining MMSE score has a great impact on AD detection.

The contributions of this study are as follows: (1) we adapt a technique to generate specific individual features from indirectly constructed brain networks based on sMRI and PET data, making it possible to use GNN to model these data modalities; (2) The association between individual features and subjects in the population is represented by combining imaging data with phenotypic data, and we discussed the effect of phenotypic information on GNN diagnostic performance; (3) To use the complementary relationship between image information ignored in graph construction, the adjacency matrices constructed by different imaging features are fused to realize edge weight sharing; (4) Through a combination of a late fusion strategy, our proposed multimodal GNN framework is further improved in AD diagnosis performance.

## 2. Dataset and Materials

The data used in this work are from the Alzheimer's Disease Neuroimage Initiative (ADNI) database [19], which is publicly available (www.loni.ucla.edu/adni). We used the MPRAGE sMRI and FDG-PET (six 5-min frames 30-60 min post injection) from the ADNI-1 and ADNI-2 baseline for AD assessment, acquiring paired multimodal images from the same subject and from the closest acquisition date. The detailed description of image protocols and acquisition can be found at adni-info.org. Except for AD and NC subjects, the obtained MCI data are divided into progressive MCI (pMCI) and stable MCI (sMCI): MCI subjects who developed AD within 3 years were classified as pMCI and those who did not convert to AD were classified as sMCI. In total, the dataset has 792 subjects, including 215 AD, 246 NC, 331 MCI (120 MCI converters (pMCI), and 211 MCI non-converters (sMCI)). The MMSE is a cognitive scale with scores ranging mainly from 10-30, with 30 indicating normal cognitive impairment and lower scores indicating more severe dementia. The gene data apoe4 in our study includes three genetic types tagged as 0, 1 and 2. Table 1 shows the key demographic statistics for each category of subjects in this study.

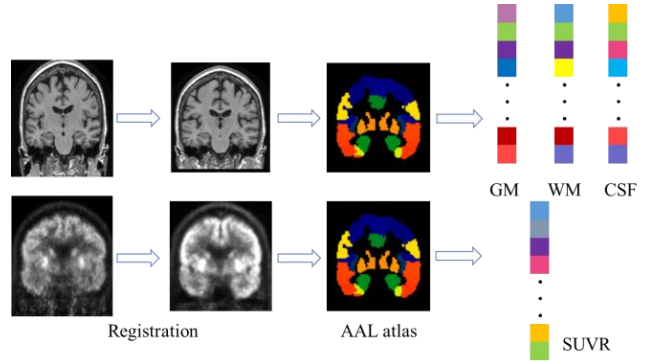

**Fig. 1.** The preprocessing pipeline of sMRI and PET scans. The raw brain images were aligned to MNI152 space and then ROI features were extracted for each modality image, using AAL atlas.

**Table 1**. The demographic information of dataset used in this study

|      | Numbers | Gender(M/F) | Age(yrs) | MMSE(score) |
| --- | --- | --- | --- | --- |
| AD   | 215 | 126/89 | 74.9±7.7 | 23.21(±2.13) |
| NC   | 246 | 125/121 | 74.1±5.8 | 29.02(±1.21) |
| sMCI | 211 | 125/86 | 72.5±7.4 | 28.01(±0.71) |
| pMCI | 120 | 74/46 | 74.4±7.1 | 27.15(±1.81) |

We used conventional procedures for brain image preprocessing, correction, and affine registration; the data preprocessing workflow is shown in Fig. 1. Specifically, all sMRI data underwent anterior commissure-posterior commissure correction and affine alignment via SPM12. The N4 algorithm [20] was applied to correct the non-uniform tissue intensities and affine alignment to MNI152 space [21] was done to align the sMRI with the normalized template. PET images were co-registered to the corresponding N4 bias-corrected sMRI images by using rigid and non-linear for co-registration routine by Clinica platform [22][23]. The resolution of processed images was 121×145×121. After that, we extracted 116 sMRI ROI features and 116 PET ROI features based on the AAL atlas [24], respectively. For sMRI, the volumetric information of gray matter (GM), white matter (WM), and cerebrospinal fluid (CSF) in brain ROI regions were obtained. For PET, the standardized uptake value ratio (suvr) [25][26] in brain ROI regions was obtained. The calculation of suvr is relative to each individual brain region. We divided the data according to the ID No. of subjects, with the first $n$-1 numbered subjects used for network training, and half of the data after $n$ used for validation and the other half for testing. To avoid data leakage [27], all brain images in each modality dataset were not from the same subject.

## 3. Methods

In this study, we propose a multi-modal GNN architecture to perform early detection of AD. The architecture is composed of multiple branches of GCN, one for each data modality. Nodes in the adjacency matrix used in the GCNs represent single modality features from a single subject. The scores of all subjects are computed through the decision-making output of a softmax layer in each branch, which are then combined for the final prediction. To better capture the relationships between subjects with image features, we propose to construct a brain network for each subject from ROI features extracted from imaging data, instead of directly using the ROI features from the brain. The edges of the adjacency matrix are defined by combining features from the brain networks constructed with the phenotypic information of subjects, which reveals the similarity between the features of each subject. This bears some similarity with the population graph approach which has become popular recently [12]. A key difference and novelty in our proposed approach is the application to multimodality of images and the method of fusing graphs generated from each data modality.

### 3.1. Individual Feature extraction

In the population graph, each node represents features of a subject. Due to slight differences between ROI features from sMRI and PET images, using these features in an input matrix leads to suboptimal model performance. Instead, we construct a brain network to extract more contrasting features so as to achieve better performance.

#### 3.1.1. Individual features based on PET

For PET ROI features, it is unclear how to construct brain networks since ROI features are in the form of a vector (unlike fMRI, which is a 4-dimensional data and it is straightforward to see how a correlation matrix can be built). Therefore, we construct a brain network [18] for every subject indirectly by comparing them to a group of normal subjects.

First, we calculate the weighting matrix based on the interregional effect size differences of average intake between individual subjects and mean NC subjects. The connectivity $E(i, j)$ of a subject in the $i$-th ROI and $j$-th ROI is expressed as:

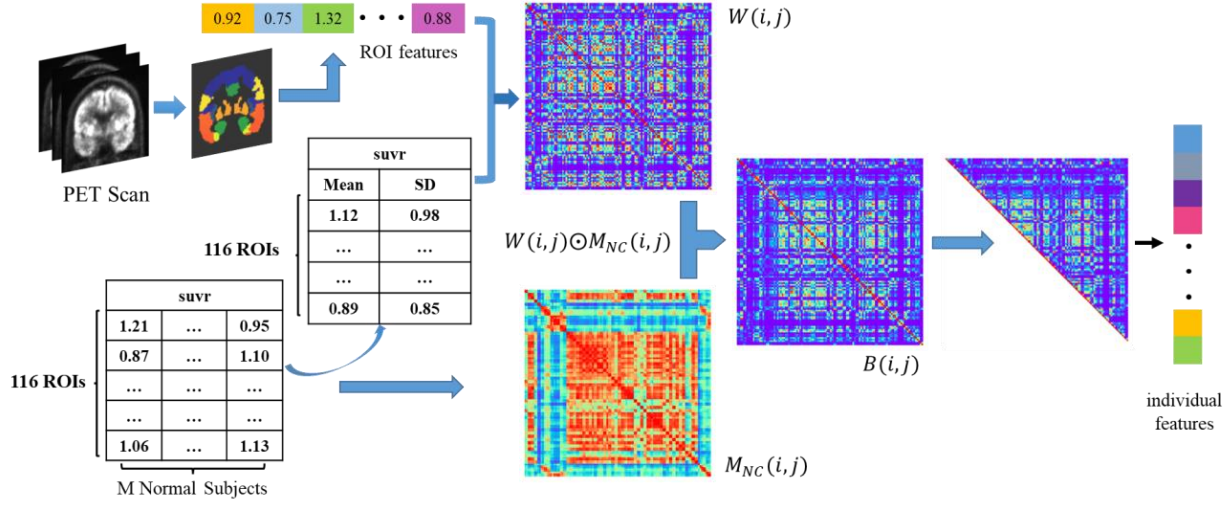

**Fig. 2.** The flow chart of individual brain network and feature extraction from PET image. We obtained the ROI features (suvr) of brain regions from PET, then derived the mean and standard deviation of ROI features based on a group of normal health subjects, and obtained the brain matrix by our computational process, and finally flattened the upper triangular matrix into a one-dimensional individual features.

$$E(i, j) = \frac{\left|(f_i - \overline{f}_{NC,i}) - (f_j - \overline{f}_{NC,j})\right|}{s_p(i, j)} \quad (1)$$

where $f_i$ represents the metabolic information suvr of a person in the $i$-th ROI; and $f_{NC,i}$ represents the average metabolic information of all NC patients in the $i$-th ROI. In formula (1), $s_p(i,j) = \sqrt{(s_i^2 + s_j^2)/2}$ where $s_i$ represents the standard deviation of the metabolic information of all NC subjects in the $i$-th ROI.

The expression of correlation coefficient value $R(i, j)$ between the $i$-th and $j$-th ROIs is obtained based on Fisher transform [28]:

$$R(i, j) = \frac{\exp(2 \times E(i, j)) - 1}{\exp(2 \times E(i, j)) + 1} \quad (2)$$

the value of $R(i, j)$ ranges between 0 and 1, and decreases with the increase of $E(i, j)$. Then, the weighting matrix $W(i, j)$ of a single subject is expressed as:

$$W(i, j) = 1 - R(i, j) \quad (3)$$

The weighting matrix $W$ of a subject is then multiplied by the connectivity matrix of the NC group to obtain the connectivity between the $i$-th ROI and the $j$-th ROI of a subject. The brain network matrix $\{B(i, j)\}$ is expressed as:

$$B(i, j) = W(i, j) \odot M_{NC}(i, j) \quad (4)$$

where $M_{NC}(i, j)$ is the value of row $i$ and column $j$ in the correlation coefficient matrix made by each ROI of all NC subjects and $\odot$ indicates Hadamard product. A flow chart of the process of creating individual brain matrix and feature extraction is shown in Fig. 2.

Finally, through the feature extraction from the subject's brain matrix $B$, we use the values on the upper triangle of matrix $B$ as the subject's individual features. Taking the subject with $P$ brain ROIs as a reference where the dimension of the connectivity matrix $B$ is $P \times P$. Then the dimension of the individual features is given by $(P \times (P+1))/2$.

### 3.1.2. Individual features based on sMRI

The ROI features obtained from sMRI images include gray matter (GM), white matter (WM), and cerebrospinal fluid (CSF). Therefore, we can construct the corresponding individual brain network separately by using several ROI features (GM, WM, or brain matter (GM+WM)) extracted from sMRI in accordance with the above method (3.1.1). Furthermore, we need to explore the specificity of different features obtained by above method to provide more effective input features for the multi-modal GNN.

### 3.2. Graph construction

The performance of GCN is greatly influenced by how its adjacency matrix is constructed [29]. In this work, each node in the graph is represented by the feature vectors of its corresponding subject, and the edge weights between nodes represent the similarities between the subjects [9][12]. We define an undirected graph $G(V, E, A)$ with a set of vertices $v_n \in V$ $(n=1,2,\dots,N)$ where $n$ represents the number of subjects. Each vertex $v_n$ is represented by the subject associated with the features from the upper half matrix of each brain network and the edges $(v_n, v_m) \in E$, $(v_n, v_m) = a_{nm} = a_{mn}$, $a_{mn} \in A$ where each element of $A$ is an edge weight. $A$ is a normalized adjacency matrix describing the connectivity of all vertices. The normalized graph Laplacian is defined as:

$L = I - A = I - D^{-1/2}WD^{-1/2}$, where $D = diag_i\left[\sum_{i \neq j} w_{ij}\right]$ is the diagonal degree matrix. Generally, we can obtain the adjacency matrix $A$ by computing the similarity. For a total of $N$ subjects, each subject is represented as a node, where each node is assigned a label $l \in \{0, 1\}$ corresponding to its class. The two-layer GCN can be described by the formula:

$$Z = \text{softmax}(A\text{ReLU}(AXW^{(0)})W^{(1)}) \quad (5)$$

### 3.2.1. Edge connections and weights

Edge connections and edge weights are key features in GCN as they determine which nodes are used to perform convolution and corresponding convolution coefficients. Edge weights are calculated in different ways in various studies [12][15]. In this work, we combined non-imaging information to construct the graph that established edge connections for assigning larger edge weights among subjects.

In graph theory, the initial similarity can be used to construct the edge weights for convolution filtering. We estimate the similarity $S$ between subjects $v$ and $u$ by calculating the correlation distance. The $S$ is defined as:

$$S(F_v, F_u) = \frac{\exp(-[\rho(F_v, F_u)]^2)}{2\sigma^2} \quad (6)$$

where $\rho$ is the correlation distance, $\sigma$ is the width of the kernel, and $F_v$ and $F_u$ are the feature vectors of the subject $v$ and subject $u$.

To this end, we further consider the non-imaging information (such as gender, gene and MMSE score, etc.) to construct an adjacency matrix $A(v, u)$, which is calculated as:

$$A(v,u) = S(F_v, F_u) \times (r_G(G_v, G_u) + r_P(P_v, P_u) + r_M(E_v, E_u)) \quad (7)$$

In formula (6), the $r_G$, $r_P$ and $r_M$ are defined as:

$$r_G(G_v, G_u) = \begin{cases} 1, & G_u = G_v \\ 0, & G_u \neq G_v \end{cases} \quad (8)$$

$$r_P(P_v, P_u) = \begin{cases} 1, & P_v = P_u \\ 0, & P_v \neq P_u \end{cases} \quad (9)$$

$$r_M(M_v, M_u) = \begin{cases} 1, & |M_u - M_v| \leq 1 \\ 0, & \text{otherwise} \end{cases} \quad (10)$$

where $r_G$ represents the distance for their gender information and $r_E$ represents the distance for their apoe4 information and $r_M$ represents the distance for their MMSE score. When the corresponding two subjects have the same gender or same apoe4 or similar MMSE socre, the edge weight is doubled, otherwise it is set to 0 as shown in formula (8), (9) and (10).

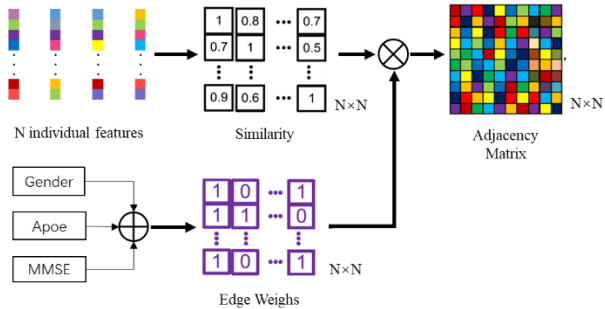

**Fig. 3.** The flow chart of the constructing of adjacency matrix combined with phenotypic information weighs.

The constructing of adjacency matrix combined with phenotypic information weights is shown in Fig. 3. The above approach of constructing the adjacency matrix $A$ works for a single modality but it does not describe how to deal with multiple modalities. We address this issue in the next two subsections.

### 3.2.2. Integration mechanism for adjacency matrices

Due to the complementarity of structural and functional information, we further construct an integrated adjacency matrix that combines the adjacency matrix from individual modalities. Based on the above construction method of adjacency matrix $A(v, u)$, we can obtain the adjacency matrix $A_s$ based on sMRI features and the adjacency matrix $A_f$ based on PET features respectively, and then the integrated adjacency matrix $A_{im}$ is calculated by Hadamard product:

$$A_{im} = A_s \odot A_f \quad (11)$$

### 3.2.3. Fusion mechanism for node vectors

According to (6), we estimate the similarity $S$ between subjects $v$ and $u$ by calculating the correlation distance between feature vectors. To fuse the two modality features to obtain a shared adjacency matrix, we concatenate the individual features of two images to calculate the correlation matrix. Then $S$ can be expressed as:

$$S(F_{vc}, F_{uc}) = \frac{\exp(-[\rho(F_{vc}, F_{uc})]^2)}{2\sigma^2} \quad (12)$$

where $F_{vc}$ and $F_{uc}$ are the concatenated feature vectors of two modality images of subject $v$ and subject $u$. Then the adjacency matrix $A_{fm}$ based on fusion mechanism can be calculated by using (7).

### 3.2.4. Integrated fusion mechanism

Through the above two mechanisms, we further construct a shared adjacency matrix to fuse the adjacency matrices of each modality. Based on the construction methods of the above two adjacency matrices $A_{im}$ and $A_{fm}$, the integrated fusion adjacency matrix $A_{if}$ is calculated by Hadamard product:

$$A_{if} = A_{im} \odot A_{fm} \quad (13)$$

### 3.3. Chebyshev GCN

In GCNs, spectral theory improves the adjacency matrix by applying Fourier transform and Taylor expansion to obtain an excellent filtering effect. The spectral domain convolution on graphs [9] can be expressed as the operation of signal $x$ with the filter $g_\theta = \text{diag}(\theta)$ by:

$$g_\theta * x = U g_\theta(\Lambda) U^T * x = \sum_{k=0}^{K} \theta_k T_k(\tilde{L}) x \quad (14)$$

where $U$ is the eigenvector matrix and calculated by the formula $L = I_N - D^{-1/2} A D^{-1/2} = U \Lambda U^T$. $I_N$ and $D$ is the identity matrix and diagonal degree matrix, respectively. The truncated expansion of Chebyshev polynomials is well approximated to $g_\theta(\Lambda)$ of $K$-order [30]. $\theta_k$ is the vector of Chebyshev coefficients, $T_k$ is the Chebyshev polynomial function, and $\tilde{L} = 2/\lambda_{max}\Lambda - I_N$. Different filtering effects can be obtained by adjusting the polynomial order $K$, the best performance is achieved when $K$ is set to 3 or 4 [12].

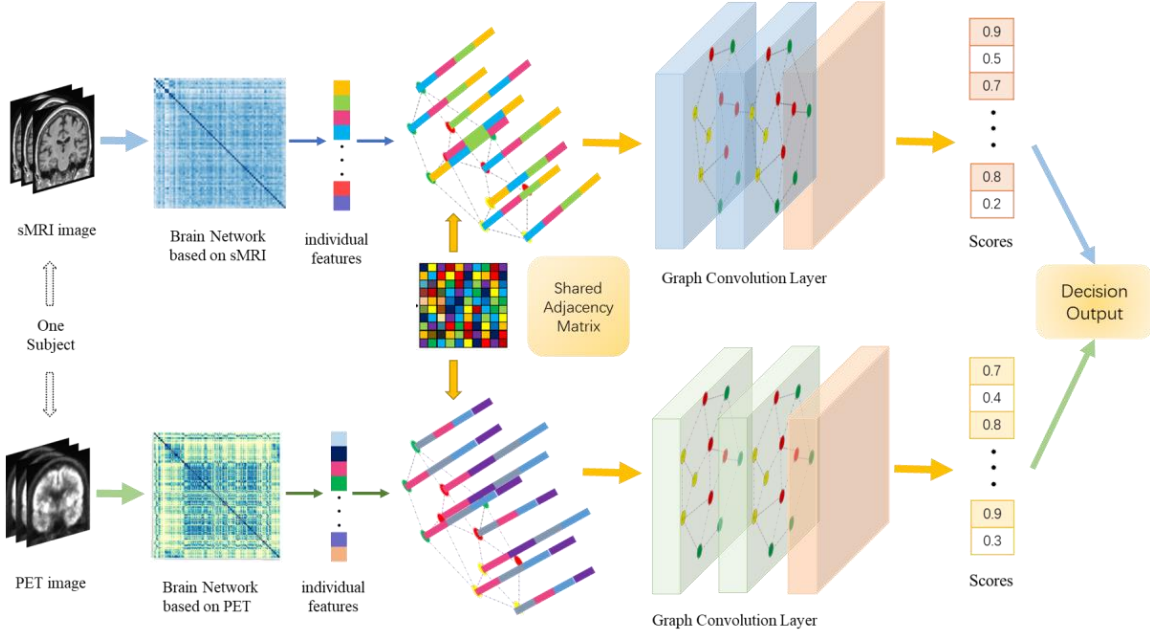

**Fig. 4.** The network architecture of our multi-modal GNN method. We incorporate phenotypic information into the graph, where the nodes represent subjects and it associates the subject's imaging features. There are two layers composed in the graph network and finally decision is derived by late fusion mechanism, each score in branch network corresponds to the diagnostic result of the corresponding subject.

*3.4. Multi-modal network architecture*

Our multi-modal network framework consists of two branches of Chebyshev GCN (CGCN), one for each modality. Each branch consists of a two-layer CGCN where hidden layers are activated by ReLU function, the number of units in hidden layer is $L$ ($L$=32). In each branch, the output layer is followed by a softmax function. The trained GNN marks the unlabeled nodes on the test set and outputs the scores by softmax. We use dropout after the ReLU activation of each layer to reduce overfitting. The softmax function for $N$ class probabilities output of the sub network is as follows:

$$\text{softmax}(z_j) = \frac{\exp(z_j)}{\sum_{j=1}^{N}(\exp(z_j))} \quad (15)$$

where $z_j$ in the above (13) represents the $j$-th value of the output vector in network. $N$ is the number of categories, the calculated softmax($z_j$) value is between (0, 1).

After the softmax function in each branch, we get the final prediction result by the decision fusion of the output probability of softmax:

$$\text{softmax}(z_j)_{final} = \frac{1}{2} \times \left( \frac{\exp(z_{j0})}{\sum_{j0=1}^{N}(\exp(z_{j0}))} + \frac{\exp(z_{j1})}{\sum_{j1=1}^{N}(\exp(z_{j1}))} \right) \quad (16)$$

where $z_{j0}$ is from the first branch and $z_{j1}$ is from the second branch.

Our multi-modal network architecture is illustrated in Fig. 4. In the population-based GNN, the training set is a labeled subset of graph nodes and the trained GNN produces classification labels for the unlabeled nodes in the test set.

## 4. Experiments and Results

*4.1. Experimental design*

Our models were implemented in PyTorch and ran on a Windows x86-64 computer equipped with Intel(R) Xeon(R) @3.60GHz, NVIDIA Quadro P620 and 32GB memory. In experiments, the training set, validation set and test set was obtained by partitioning the dataset of proportions 70%, 15% and 15%. In our population-based GNN, the training set and verification set were labeled while the test set was unlabeled with a mask. The labels of the test set are unknown during the network optimization, the test set is predicted after training and compared with the correct labels to derive the performance measures. Due to the limited access to medical images compared to other fields, especially in the case of multi-modality data for the same subject, current studies [16] [27] are mainly based on ADNI, the most world widely used database. However, the ADNI does not specify a fixed test set. To show that our method still has some generalization, our experiments were conducted on four different sub-datasets (the test set of four sub-datasets do not include the same subjects) and calculated the average accuracy to report as the final diagnostic result. Further, we selected two sub-datasets to evaluate the stability of model by a five-fold cross-validation strategy.

The hyperparameters were determined empirically as follows: dropout rate was 0.5, weight decay was set to 5e-4, and learning rate was set to 1e-3. The order $K$ in CGCN was set to 3. The network was trained for 100 epochs for convergence, and we compared with the GCN architecture trained for 300 epochs. The cross-entropy loss function was used to optimize the model parameters. To compare the validity of proposed method, the training hyperparameters were fixed in all methods. In addition to the AD vs. NC classification for disease exclusion, the prediction of MCI conversion is of great importance for the early treatment of AD patients. Therefore, we conducted the classification tasks of AD vs. NC and sMCI vs. pMCI, and evaluated the performance based on accuracy (ACC), sensitivity (SEN), specificity (SPE) and the area under curve (AUC).

We divided the experimental section into the following parts. Firstly, we carried out experiments on GCN model based on single modality images (i.e., separate experiments for sMRI and PET), and compared the diagnostic effectiveness of brain network features constructed based on several brain features. Secondly, we experimented with several multi-modal methods and compared it with the single modality approach. Finally, we carried out the explorative experiments by constructing the adjacency matrix combined with gender, apoe4, MMSE, etc. information. Using phenotypic information can further improved the diagnostic performance of GNN, and discussed the impact of phenotypic information on AD diagnosis. In addition, we compared the state-of-the-art method to prove the effectiveness of our proposed method.

### 4.2. Experimental results and discussion

First, the GCN model based on sMRI features was used for ablation experiments and the impact of sMRI features for AD diagnosis was explored. Features obtained through brain networks (BN) and directly extracted ROI features were used as inputs for GCN in this part. We know that in sMRI, cognitive impairment is mainly related to atrophy of GM, WM and brain structure [31][32]. For this reason, first we obtained individual features in this way (Section 3.1.2) based on three kinds of ROI features (GM, WM and brain matter) separately, as well as the way [12] of ROI features extracted directly from MRI, then carried out classification experiments based on GCN model. Results from these experiments are summarized in Table 2. Compared with the GM ROI features from MRI, the individual features we obtained through the brain network have better specificity, allowing the GCN model to perform better clustering performance with a considerable improvement in accuracy. Moreover, from the experimental results, it is seen that models using GM features produced the best result in AD vs. NC classification, with an average accuracy of 87.71%. The models using brain matter features produced the best result in sMCI vs. pMCI classification, with an average accuracy of 72.40%. This also reflected that in AD symptoms, the biomarkers of gray matter are more specific, while in MCI period, the atrophy of gray matter is not obvious compared with AD.

Secondly, we carried out the similar experiments based on PET metabolic features. The results of ablation GCN

**Table 2.** The classification results of GCN model based on various imaging information from sMRI and PET

| Features | AD vs. NC | | | | sMCI vs. pMCI | | | |
|---|---|---|---|---|---|---|---|---|
| | ACC | SEN | SPE | AUC | ACC | SEN | SPE | AUC |
| GM ROIs | 81.62 | 78.74 | 83.83 | 81.28 | 65.00 | 31.18 | 85.99 | 59.42 |
| GM+WM BN | 85.15 | 81.17 | 88.51 | 84.84 | 72.40 | 48.15 | 87.47 | 67.81 |
| WM BN | 77.14 | 69.56 | 83.62 | 76.59 | 67.20 | 36.74 | 87.50 | 62.12 |
| GM BN | 87.71 | 83.95 | 91.43 | 87.63 | 71.20 | 41.33 | 88.40 | 64.87 |
| PET ROIs | 82.45 | 74.22 | 87.81 | 82.02 | 66.23 | 42.22 | 81.56 | 61.89 |
| PET BN | 88.00 | 86.54 | 89.37 | 87.95 | 73.20 | 53.26 | 82.96 | 68.06 |

BN (Brain Networks)

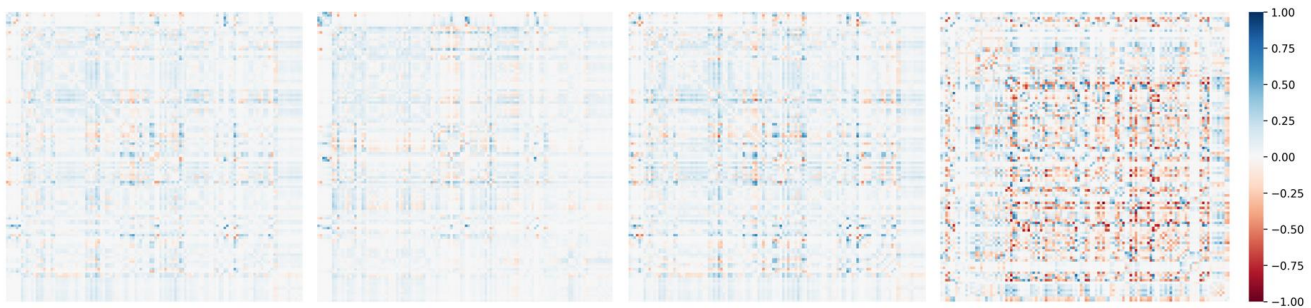

**Fig. 5.** The four figures are the difference of individual brain network matrix between AD and NC subjects. The matrices from left to right are based on brain matter features, WM features, GM features and PET features respectively.

experiments based on PET ROI features and the features from the brain network (Section 3.1.1) are also shown in Table 2. From the classification results in Table 2 based on sMRI and PET features, the method of constructing individual brain network of PET features has better performance. Meanwhile, the accuracy of diagnosis based on brain network features is much higher than that based on ROI features. Brain ROI feature methods are often based on traditional machine learning such as support vector machines (SVM), which needs to achieve better performance with effective feature selection [35][36], but this also requires more processes and is usually effective on smaller samples. Our features acquisition from brain network shows better advantages in terms of performance and efficiency.

Furthermore, the value of using GM features can be demonstrated by visualizing and comparing the brain matrices built using various imaging features. As seen in Fig. 5, the difference between AD and NC is the greatest for GM amongst structural image features. Also, PET shows an obvious difference between AD and NC that were even larger than those seen in GM. This might explain why models using PET did better than models that used sMRI features. This result is consistent with the established clinical knowledge. PET can detect the functional brain changes and specific pathologies of AD at the early stage than sMRI.

By the correlation coefficients of brain regions in our constructed brain network based on sMRI features and PET features in AD diagnosis, we selected five key regions. Fig. 6 shows the visualization of the key ROIs in brain for AD diagnosis based on sMRI and PET in our study. In sMRI, specifically, these ROIs are Temporal_Pole_Sup, Rectus, Lingual, Hippocampus and Amygdala. In PET, specifically, these ROIs are Frontal_Sup, Frontal_Sup_Medial, Occipital_Mid, Occipital_Inf and Temporal_Mid. It can be seen that some of these brain regions are mainly concentrated

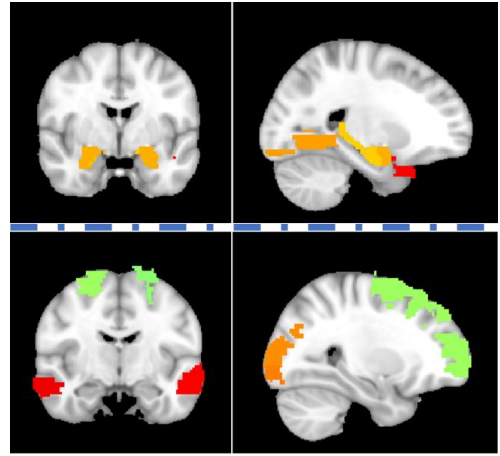

**Fig.6**. Visualization of the key ROIs in brain for AD diagnosis. In the top row we show the key ROIs in the coronal and sagittal views of brain MRI image. In the bottom row we show the key ROIs in the coronal and sagittal views of brain PET image.

in memory regions, which are correlated with cognitive disorders in some clinical studies [35][36].

In the subsequent experiments on multi-modal datasets, we will therefore focus on models that use GM features as the structural imaging modality. The similarity only is used to construct the adjacency matrix and we use both types of GNN models for comparative tests, including GCN and CGCN. To test whether the combination of multi-modal imaging features can improve diagnostic performance, we experimented with several multi-modal mechanisms in this study. We first create a baseline where the two GCN branches are simply combined, which we call dual GCN (DGCN), that is, each branch uses its own adjacency matrix. Then, we designed different multi-modal fusion techniques that constructs a shared adjacency matrix in three different ways: integration DGCN (IDGCN) from Section 3.2.2, fusion DGCN (FDGCN) from Section 3.2.3 and integrated

**Table 3.** The classification results of several multi-modal methods based on GCN model

| Methods | AD vs. NC | | | | sMCI vs. pMCI | | | |
|---|---|---|---|---|---|---|---|---|
| | ACC | SEN | SPE | AUC | ACC | SEN | SPE | AUC |
| DGCN | 89.65 | 87.94 | 91.13 | 89.53 | 73.50 | 51.83 | 85.99 | 68.91 |
| IDGCN | 90.36 | 86.47 | 93.87 | 90.17 | 74.00 | 51.03 | 87.78 | 68.67 |
| FDGCN | 90.71 | 87.94 | 93.12 | 90.55 | 75.00 | 52.05 | 85.34 | 68.69 |
| IFDGCN | 91.07 | 88.72 | 93.25 | 90.98 | 75.50 | 50.25 | 88.13 | 68.45 |

The methods in this table are based on GCN, DGCN means Dual GCN, IDGCN means Integration Dual GCN, FDGCN means Fusion Dual GCN, IFDGCN means Integrated Fusion Dual GCN.

**Table 4.** The classification results of several multi-modal methods based on Chebyshev GCN model

| Methods | AD vs. NC | | | | sMCI vs. pMCI | | | |
|---|---|---|---|---|---|---|---|---|
| | ACC | SEN | SPE | AUC | ACC | SEN | SPE | AUC |
| DCGCN | 90.00 | 88.60 | 91.12 | 89.86 | 74.50 | 48.29 | 87.67 | 65.01 |
| IDCGCN | 90.72 | 89.41 | 91.86 | 90.63 | 75.00 | 49.43 | 88.02 | 68.72 |
| FDCGCN | 91.07 | 90.15 | 91.86 | 91.00 | 75.50 | 50.43 | 88.02 | 69.22 |
| IFDCGCN | 91.07 | 90.22 | 91.87 | 91.04 | 75.50 | 49.90 | 88.70 | 69.30 |

The methods in this table are based on CGCN (Chebyshev GCN), DCGCN means Dual CGCN, IDCGCN means Integration Dual CGCN, FDCGCN means Fusion Dual CGCN, IFDCGCN means Integrated Fusion Dual CGCN.

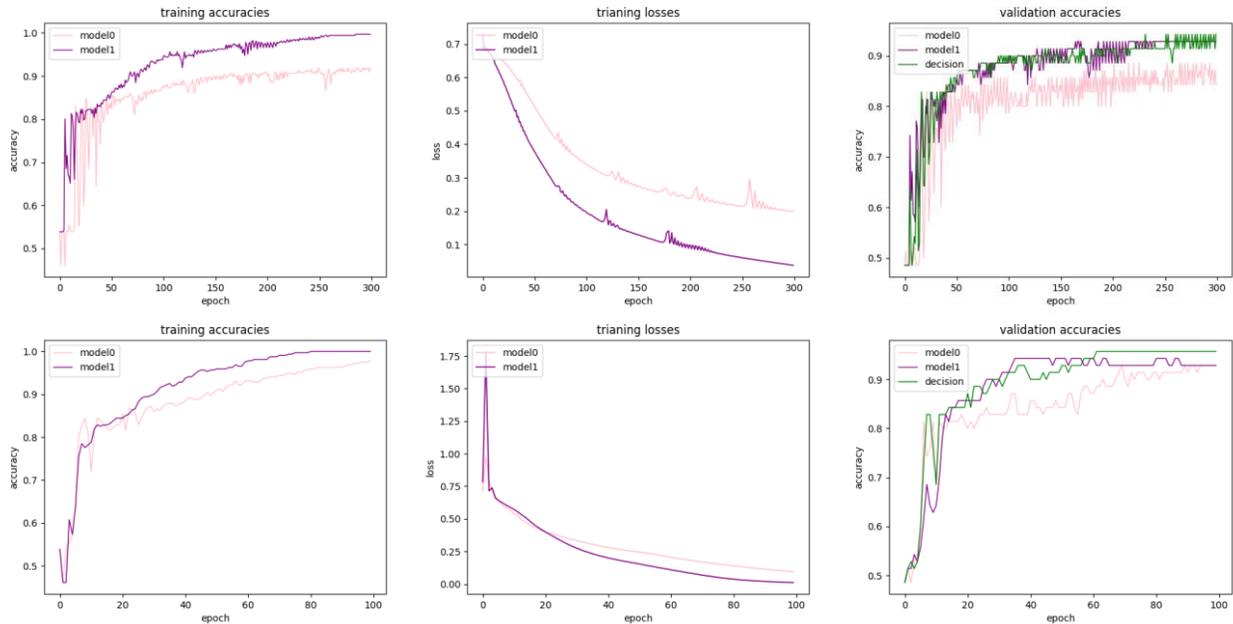

**Fig.7.** The figures from left-to-right are the training curve, loss curve and verification curve based on IFDGCN (above) and IFDCGCN (below) multi-modal method, respectively.

fusion DGCN (IFDGCN) from Section 3.2.4. The results based on GCN are shown in Table 3, and the results based on CGCN are shown in Table 4. In the binary classification of AD vs. NC, IFDCGCN achieves the best accuracy of 91.07, sensitivity of 90.22, specificity of 91.87 and AUC of 91.04. In the binary classification of sMCI vs. pMCI, IFDCGCN achieves the best accuracy of 75.50, and its corresponding sensitivity of 49.90, specificity of 88.70 and AUC of 69.30 are also improved as compared to single modality methods.

In the population-based GNN method for AD diagnosis, the effective expression of individual features can lead to better prediction performance. From the results in Table 3 and Table 4, we demonstrated that our proposed multi-modal fusion framework can further improve the accuracy of AD diagnosis. The effectiveness of our multi-modal method can be attributed to the following three points. Firstly, it is evident that the late fusion mechanism helped to improve the accuracy of the model prediction. The late fusion combines the decisions of two independent branches of GNN, which is consistently observed in both GCN and CGCN models. CGCN performs better in accuracy as compared to GCN. Also, CGCN has the advantage of stability as the standard deviation of its results is smaller. In Fig. 7, we showed the training curves, loss curves and validation curves of the IFDGCN and IFDCGCN multi-modal methods for AD prediction. Seen from the validation curves, the accuracy of the fusion decision is higher than that of the two separate branches when the network training reaches a certain epoch. According to the training, validation curves and epoch, CGCN converges faster, and the prediction is more stable. Secondly, the multi-modal mechanisms we proposed are more effective than the simple late fusion approach in DGCN.

This shows the value of creating a shared adjacency matrix constructed based on multi-modal data. The way adjacency matrices are constructed has a direct impact on the performance of the GNN models.

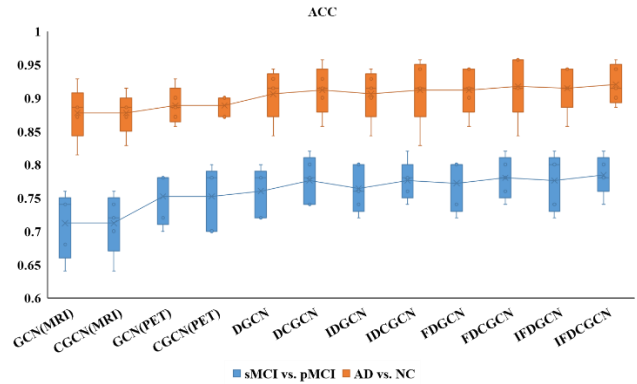

**Fig. 8.** The accuracy of classification results based on single modality and multi-modal methods.

Fig. 8 summarizes the comparisons between single modality methods and multi-modal methods in a box plot showing classification accuracy. While multi-modal methods are clearly superior to single modality approaches, we note that the choice of integration mechanisms does not lead to huge differences in model performance. In addition, the multi-modal method of CGCN has better performance in the choice of GNN models, and the stability of CGCN is much better than that of GCN. Overall, our proposed IFDCGCN produced the best results in terms of classification performance and stability.

**Table 5.** The classification results of Multi-modal GNN framework combining the phenotypic information

| sMRI+PET | AD vs. NC | | | | sMCI vs. pMCI | | | |
|---|---|---|---|---|---|---|---|---|
| | ACC | SEN | SPE | AUC | ACC | SEN | SPE | AUC |
| Similarity | 91.07 | 90.22 | 91.87 | 91.04 | 75.50 | 49.90 | 88.70 | 69.30 |
| Aope4 | 91.07 | 88.56 | 93.17 | 90.86 | 75.50 | 50.03 | 88.80 | 69.42 |
| Age | 88.93 | 86.43 | 91.08 | 88.76 | 74.50 | 48.06 | 88.80 | 68.93 |
| Gender | 91.79 | 90.15 | 93.19 | 91.67 | 76.50 | 51.81 | **89.70** | 70.76 |
| MMSE | **96.68** | **99.19** | 94.49 | **96.84** | 76.00 | 51.03 | 88.80 | 69.92 |
| G+M | 95.00 | 93.09 | 96.71 | 94.90 | 77.00 | 51.90 | 89.37 | 70.63 |
| G+A+M | 93.21 | 90.19 | **95.98** | 93.08 | **78.00** | **54.96** | 89.37 | **72.16** |

G+M means the combining of gender and MMSE. G+A+M means the combining of gender, apoe4 and MMSE.

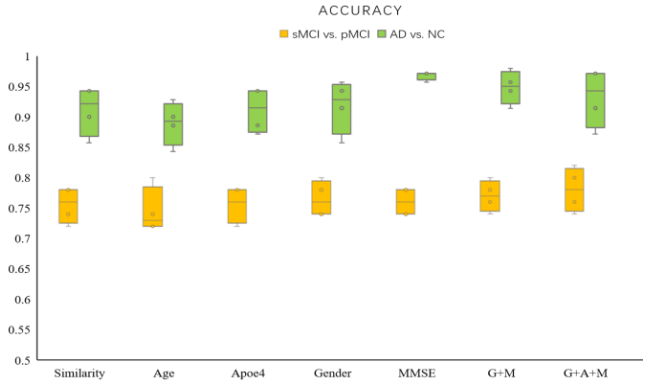

**Fig. 9.** It summarizes the comparison of diagnostic accuracy based on multi-modal GNN combined with various phenotypic information.

In addition, some non-imaging information has also been found to be associated with AD in some studies [37][38][39], such as genes, gender, age, MMSE, etc., which have important references in the diagnosis of cognitive impairment. With the accumulation and richness of imaging and non-imaging multi-source data, how to fuse multi-source and multimodal data is the trend for accurate AD evaluation in the future. Therefore, this work aims to further fuse non-imaging information based on the use of multi-modal GNN framework to achieve more accurate AD diagnosis.

The results of the ablation experiments on AD vs. NC and sMCI vs. pMCI diagnostic tasks were further explored based on our validated multi-modal GNN (IFDCGCN) incorporating various phenotypic information as shown in Table 5, and the Fig. 9 summarizes the corresponding comparisons in a box plot. We found that the adjacency matrix combined with gender, gene, and MMSE score information benefited or improved the diagnostic performance of the model, especially combining MMSE (in formula (7), $r_G=0$ and $r_P=0$) obtained a very significant improvement in AD vs. NC diagnosis. The best results were obtained by MMSE, gender, and gene all weighted information in sMCI vs. pMCI diagnosis. With the information based on MMSE scores, it made the weighting more pronounced in AD vs. NC subjects, while both sMCI and pMCI belong to MCI patients, so their MMSE scores were close, both almost in the range of 27-29 score, this made the inter-subject weights insignificant to the extent that the improvement in GNN classification performance is limited. But also obtained better prediction of MCI conversion with the combining of several phenotypic information. In contrast, combining age information (age difference within 1 year weighted as 1, otherwise 0) in our GNN approach did not have any improvement or even a decrease in disease prediction. In the graph, the weights of non-imaging information are associated with the imaging features, the combination of effective phenotypic information allows to target a few subjects with marginal imaging features to be judged correctly. In this part of the explorative experiments, e.g., gender information also plays a role in the construction of the adjacency matrix on the performance of the GCN, which is consistent with some results of study [12]. Some phenotypic information is more clinically accessible, which has an advantage for GNN-based AD diagnosis methods. For the age information, the effect is not ideal. We infer that it is difficult to find a direct correlation with the features of subjects because of the wide range of age distribution. However, age information is helpful for the diagnosis of AD in clinical practice, which is also what needs further research in future.

On a fixed two sub-test set, our IFDCGCN method combining the MMSE information was again experimented with 5-fold cross-validation in the tasks of AD vs. NC, and the results shown in Fig. 10. The average accuracy was 98.00% and 96.29% with standard deviations of 0.78 and 0.78, respectively. The AUCs were 98.06 and 96.58 with standard deviations of 0.90 and 0.72, respectively. The above results indicate that our proposed multi-modal GNN is stable.

In this work, the adjacency matrix can be constructed based on a combination of similarity matrix and non-imaging data. To better demonstrate the advantages of multi-modal mechanisms and to explore the differences in constructing adjacency matrices based on phenotypic information, Fig. 11 compares the visualizations of the adjacency matrices constructed using MMSE and GAM (Gender+Apoe4+MMSE) information based on IFDCGCN method in two diagnostic tasks. In these visualizations, we rearranged the rows so that subjects in the same category were grouped together to make the differences between categories more apparent. The group similarity matrices constructed based on our integrated fusion mechanism in combination with MMSE showed a very significant intra-group correlation for the AD vs. NC subject group, resulting in an average accuracy of 96.68%. In contrast, the combination based on multiple phenotypic information had relatively better intra-group correlations in sMCI vs. pMCI diagnosis. However, MCI conversion prediction needs to be continuously explored, and since both sMCI and pMCI belong to the MCI category, the low sensitive features of both types also contribute to the lower prediction accuracy. We infer that it is more important to acquire or construct individual features that are more perceptive.

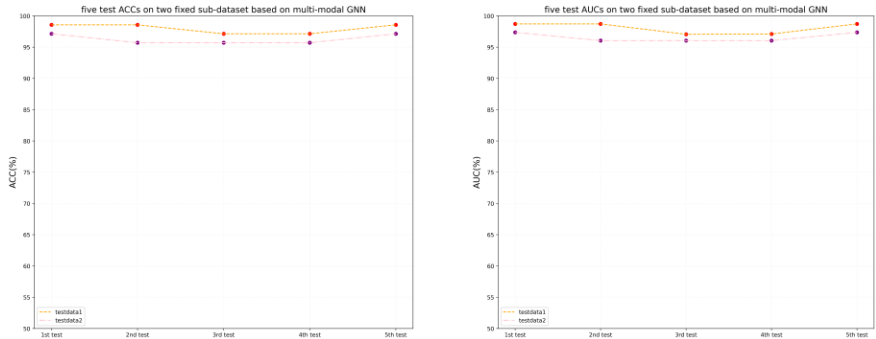

**Fig. 10.** The five test ACC and AUC results by 5-fold cross-validation based on IFDCGCN multi-modal GNN on two sub-datasets, respectively.

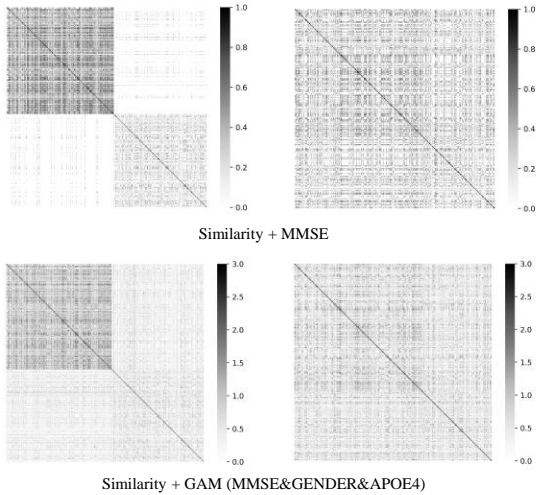

**Fig. 11.** It shows the adjacency matrices combining non-imaging data based on the IFDCGCN in AD detection (left column) and MCI prediction (right column). Our method has significant intra-group correlation and provides obvious contrast between the classes especially combining the MMSE information (top row) in AD detection (left column).

In addition to analyzing the parameters that affect the prediction performance of GNN, we also compared it with several different state-of-the-art methods based on ADNI database to verify the utility of our proposed method. The comparative studies are based on sMRI, PET and multi-modal methods. The results include AD vs. NC classification in Table 6 and sMCI vs. pMCI classification in Table 7. It can be observed that our proposed method has achieved satisfactory performance. In addition to a better prediction accuracy, it also has an advantage or comparable performance in terms of diagnostic specificity. Another notable point is our method also outperforms some multi-modal CNN methods.

In summary, the changes of brain structure and metabolic characteristics of AD patients are different, which makes multi-modal images provide more complementary information. But existing GNN analysis based on multi-modal image features are mostly limited to DTI and fMRI [16][17]: it is clear how to present these 4-dimensional data as brain networks and construct the topology of nodes for them in GNN analysis because the brain regions in fMRI or DTI imaging have the characteristics of sequential signals or fiber connection directions. However, it is not obvious how GNN can be used on sMRI and PET data. In addition, many research based on GNN methods focus on the improvement of network architecture and the optimization of adjacency matrix, while ignoring the importance of individual features. To solve the above shortcomings, we obtain specific individual features via constructing brain networks with ROI features respectively, and then construct GNN with the method of nodes representing subjects, which solves the problem of difficulty in constructing graph neural networks based on sMRI and PET features.

In our approach, we further play the advantages of multi-modal data information and improved diagnostic performance was achieved through the combination of multi-modal features, multi-modal adjacency matrices and late decision fusion. Compared with fMRI and DTI data, the preprocessing process of sMRI and PET is relatively simpler, while GNN has the advantage of being fast, flexible and more parameter efficient as compared to CNN, and easier to integrate multi-source and multi-modal data. Therefore, our work could have considerable application prospects in the task of early diagnosis of AD.

## 5. Conclusion

In this study, we proposed a population-based and multi-modal GNN to predict early Alzheimer's disease using image features and phenotypic information. Our method obtained specific individual features by constructing brain networks and combined imaging data with phenotypic data to represent the data association between individual features and subjects in potential populations. In addition, we further combined it with shared adjacency matrix and decision-making mechanism to achieve better multi-modal GNN diagnosis performance. Through several experiments, our proposed multi-modal method achieves improved prediction results on ADNI datasets especially in AD detection. Compared with several state-of-the-art methods, our proposed method shows better or equivalent diagnostic performance, including in the relatively challenging sMCI versus pMCI prediction task. Our study was mainly explorative on using ADNI dataset and further validation may be necessary using additional datasets to confirm the findings.

Table 6. The AD vs. NC classification in studies based on the ADNI dataset

| STUDY | ACC | SEN | SPE | SUBJECT NUM. | METHOD |
| --- | --- | --- | --- | --- | --- |
| Tong et al. [40] | 90.00 | 84.90 | 92.60 | 429 subjects | Graph kernels |
| Wen et al. [27] | 89.00 | - | - | 666 subjects | ROI 3D CNN |
| Wee et al. [41] | 85.80 | 83.50 | 87.50 | 1610 subjects | GCN |
| Parisa et al. [7] | 89.10 | 87.40 | 92.10 | 407 subjects | Multi-modal CNN |
| Fan et al. [42] | 88.31 | 91.40 | 84.42 | 428 subjects | Multi-modal CNN |
| Pan et al. [43] | 93.58 | 91.52 | 95.22 | 857 subjects | Multi-modal CNN |
| Lin et al. [44] | 89.26 | 82.69 | 96.48 | 670 subjects | Multi-modal CNN |
| Huang et al. [45] | 90.10 | 90.85 | 89.21 | 1378 subjects | Multi-modal CNN |
| OURS | 96.68 | 99.19 | 94.49 | 461 subjects | Multi-modal GNN |

Table 7. The sMCI vs. pMCI classification in studies based on the ADNI dataset

| STUDY | ACC | SEN | SPE | SUBJECT NUM. | METHOD |
| --- | --- | --- | --- | --- | --- |
| Wen et al. [27] | 74.00 | - | - | 593 subjects | ROI 3D CNN |
| Tong et al. [40] | 70.40 | 67.00 | 73.00 | 405 subjects | Graph kernels |
| Kang et al. [46] | 66.70 | - | - | 319 subjects | 2D CNN |
| Hett et al. [47] | 76.00 | - | - | 216 subjects | GCN |
| Parisa et al. [7] | 68.20 | 78.10 | 57.50 | 489 subjects | Multi-modal CNN |
| Lu et al. [48] | 75.44 | 77.27 | 76.19 | 626 subjects | Multi-modal Network |
| Huang et al. [45] | 76.90 | 68.15 | 83.93 | 767 subjects | Multi-modal CNN |
| Lin et al. [44] | 74.10 | 75.00 | 73.08 | 416 subjects | Multi-modal CNN |
| OURS | 78.00 | 54.96 | 89.37 | 331 subjects | Multi-modal GNN |

## Acknowledgments

Data used in preparation of this article were obtained from the Alzheimer's Disease Neuroimaging Initiative (ADNI) database. As such, the investigators within the ADNI contributed to the design and implementation of ADNI and/or provided data but did not participate in analysis or writing of this report. A complete listing of ADNI investigators can be found at: http://adni.loni.usc.edu/wp-content/uploads/how_to_apply/ADNI_Acknowledgement_List.pdf. More details can be found at adni.loni.usc.edu.

This work was partly supported by the Chengdu Major Technology Application Demonstration Project (Grant No. 2019-YF09-00120-SN), the Key Research and Development Program of Sichuan Province (Grant No. 2022YFS0098), the China Scholarship Council (Grant No. 202106240177). This work was also partly supported by AcRF Tier-2 grant 2EP20121-003 by Ministry of Education, Singapore.

## Declaration of competing interest

The authors declare that they have no known competing financial interests or personal relationships that could have appeared to influence the work reported in this paper.